\documentstyle[graphicx]{mn}

\title{The messy environment of Mrk 6}

\author[N.J.\,Schurch, R.E.\,Griffiths \& R.S.\,Warwick]
{N.J.\,Schurch$^{1*}$, R.E.\,Griffiths$^{2}$ \& R.S.\,Warwick$^{3}$\\
$^{1}$Department of Physics, Durham University, South Road, Durham, DH1 3LE, UK \\
$^{2}$Department of Physics, Carnegie Mellon University, 5000 Forbes Avenue, Pittsburgh, PA 15213, USA \\
$^{3}$Department of Physics and Astronomy, University of Leicester, University Road, Leicester, LE1 7RH, UK \\
$^{*}$ nicholas.schurch@durham.ac.uk}
\date{}

\def\gin{{\it Ginga\/}}

\def\ros{{\it ROSAT\/}}
\def\asca{{\it ASCA\/}}

\def\xmm{{\it XMM-Newton\/}}

\def\hst{{\it HST\/}}

\def\bep{{\it BeppoSAX\/}}
\def\sax{{\it BeppoSAX\/}}
\def\int{{\it INTEGRAL\/}}

\def\Msun{\hbox{$\rm ~M_{\odot}$}}

\def\H0{{\rm ~km~s^{-1}~Mpc^{-1}}}

\def\xspnorm{{\rm ~photon~cm^{-2}~s^{-1}~keV^{-1}}}

\def\arcm{{\hbox{$^\prime$}}}
\def\arcs{{\hbox{$^{\prime\prime}$}}}

\def\etal{et al.~\/}

\def\la{\mathrel{\hbox{\rlap{\hbox{\lower4pt\hbox{$\sim$}}}{\raise2pt\hbox{$<$}}}}}
\def\ga{\mathrel{\hbox{\rlap{\hbox{\lower4pt\hbox{$\sim$}}}{\raise2pt\hbox{$>$}}}}}
\def\ls{\mathrel{\hbox{\rlap{\hbox{\lower4pt\hbox{$\sim$}}}\hbox{$<$}}}}
\def\gs{\mathrel{\hbox{\rlap{\hbox{\lower4pt\hbox{$\sim$}}}\hbox{$>$}}}}

\def\d25{D$_{\rm 25}$}

\def\.25{0.25 keV\thinspace}

\begin{document}

\maketitle 

\begin{abstract}

In recent years it has become clear that understanding the absorption present in AGN is essential given its bearing on unification models.  We present the most recent \xmm ~observation of Mrk 6, with the goal of understanding the nature and origin of the complex absorption intrinsic to this source. X-ray spectral fitting shows that a simple warm absorption model provides an equally good statistical representation of the CCD data as a partial covering model. Furthermore, once the RGS data are included in the spectral fitting, the simple warm absorber model provides a very good fit to the data, without increasing the complexity of the model, in contrast with the partial covering model which requires the addition of either a low metalicity ($<$0.03 solar) thermal plasma or low temperature blackbody emission in order to provide a similar quality fit. The warm absorber is also a considerably more natural way to explain the variability observed in the X-ray absorbing column density between the previous \xmm ~observation and this one, requiring only a second, higher column density, higher ionisation, absorber to be present during the previous \xmm ~observation. In comparison, the partial covering models which requires moving, clumpy, material relatively close to the source that result in two distinct lines of sight, with separate absorbing columns that each vary considerably without any associated change in their covering fractions, in order to explain the observed variability. We associate the warm absorber either with an accretion disk wind with densities of $\sim$10$^{9}$ cm$^{-3}$, or with an ionised `skin' or atmosphere of the molecular torus with densities of $\sim$10$^{3\to5}$ cm$^{-3}$.

\end{abstract}


\section{Introduction}
\label{1}

Absorption has been widely recognised as playing a large role in shaping the X-ray spectra of active galactic nuclei (AGN), from the detailed ionised absorption features often observed in the soft X-ray spectrum of Seyfert 1 galaxies ({\it e.g}. NGC 3783; Kaspi \etal 2002, MGC-6-30-15; Fabian \etal 2002 ) through to the extreme cutoff of the X-ray continuum emission observed in Compton thick Seyfert 2's ({\it e.g.} NGC 1068; Bianchi \etal 2001, NGC 4945; Done \etal 2003). For any individual AGN, a detailed understanding of the properties of the absorption intrinsic to the source is essential to the understanding of the broad-band X-ray spectrum and, in particular, the underlying X-ray continuum emission. More generally, detailed information on the absorption intrinsic to a wide population of AGN gives us direct information on the general environment surrounding AGN and can test the geometry intrinsic to the standard unification models (Antonucci 1993, Risaliti \etal 2002). Observations with older X-ray spectroscopy satellites ({\it i.e.} \gin, \ros, \sax ~etc) supported the implied synergy between the optical classifications and the level of X-ray absorption observed in these sources ({\it e.g.} Nandra \& Pounds 1994, Smith \& Done 1996). Recent work, however, has cast considerable doubt upon the validity of the link between optical type and X-ray absorption, providing the first indications that the simplistic AGN unification model may be significantly incorrect. In particular, the identification of optical type 1 sources with considerable X-ray absorption columns ({\it e.g.} Page \etal 2001, Akiyama \etal 2000), and the converse identification of optical type 2 objects with no evidence of significant X-ray absorption ({\it e.g.} Pappa \etal 2001, Barcons \etal 2003), appear to be explicit contradictions of the implications of the standard unification model. 

It is possible to reconcile these apparent contradictions with the standard unification models but such explanations typically invoke a specific, and often complex, absorption situations ({\it e.g.} Maiolino \etal 2001). Typically, the heavy, neutral absorption observed in type-2 AGN is comparatively simple; it is well represented by absorption in single medium composed of cold, neutral, material of varying opacity from N$_{H}$$\sim$10$^{21}$-10$^{25+}$. In contrast, some objects show considerably more complex absorption properties (both in terms of profile and variability).

Studies of the physical properties of the `intermediate' class of AGN, defined by Osterbrock (1981) as having weak composite HI line profiles ({\it i.e.} having both a weak broad and a narrow component) in their optical spectra, are intriguing because of their potential to test, and impact on, the standard unification models. The classification of an intermediate Seyfert galaxy ({\it i.e.} Seyfert 1.5s, 1.8s and 1.9s) is based on the strength of the broad component of the H$\beta$ emission line relative to the broad component of the H$\alpha$ emission line in the sense that an optical spectrum with relatively prominent broad H$\beta$ is defined as a Seyfert 1.5, an optical spectrum with weak, but detectable, broad H$\beta$ is defined as a Seyfert 1.8 and an optical spectrum with no obvious broad H$\beta$ emission is classified as a Seyfert 1.9 (Osterbrock 1981, Cohen 1983). Typically, the ratio of the narrow components of the H$\beta$ \& H$\alpha$ emission can be interpreted as a measure of the reddening of the optical line spectrum, corresponding to an increase in the amount of absorbing material along the line of sight with increasing numerical sub-class (Ward \etal 1987). The ratio of the broad components, by comparison, is harder to interpret and, although reddening is one possibility, it is far from universal (see {\it e.g.} Corral \etal 2005).

In the context of the standard AGN unification models, these line properties suggest a highly restricted range of possible viewing angles. The viewing angle must be sufficiently highly inclined that the high-column density of cold material that forms the putative dusty molecular torus does not completely obstruct the innermost regions of the AGN responsible for the broad optical emission lines, yet the viewing angle must be small enough that there is a sufficient amount of material along the line of sight to significantly redden the broad component of the H$\beta$ line with respect to the H$\alpha$ line. Such a unique `glancing' viewing angle highlights that the intermediate Seyferts are the best, and possibly the only, sources that allow us to study the high column density, relatively low ionisation, absorbing material that may form a `skin' on the molecular torus or a wind from either the torus or the material immediately interior to it. The range of properties of the X-ray absorption intrinsic to intermediate Seyferts has proved to be considerable, both in terms of the complexity of the model required to model the absorption and in terms of the range of observed column densities ($\la$10$^{23+}$, Barcons \etal 2003, Schurch \& Warwick 2002). Typically the absorption is represented either by a `partial covering' model (a model in which slightly different lines of sight to the central continuum region sample separate absorption regions; {\it e.g.} Weaver \etal 1994, Feldmeier \etal 1999) or by a `warm absorber' model (a model in which the absorbing material is partially ionised; {\it e.g.} Matt \etal 2004, Schurch \& Warwick 2002). 

The Seyfert 1.5 galaxy Markarian 6 (Mrk 6), although well studied at longer wavelengths, was not extensively studied in the X-ray band prior to the advent of \xmm. Mrk 6 is an S0 spiral galaxy at a distance of $\sim$50-80 Mpc (Stark \etal 1992; here we use a distance of 70 Mpc) and was first identified as an intermediate Seyfert through optical spectroscopy by Osterbrock \& Koski (1976). A combination of near infrared and optical data were used by Rix \etal (1990) to constrain the extinction, and by association, the column density to be A$_{v}$$\ls$2 and N$_{H}$$\sim$4$\times$10$^{21}$ cm$^{-2}$ respectively. Mrk 6 is one of the few Seyferts to show considerable evidence for an ionisation cone (Meaburn \etal 1989) and there is some tentative evidence that the anisotropy of the ionising radiation is variable (Sergeev \etal 1999). MERLIN observations of Mrk 6 reveal collimated jets extending out of the plane of the galaxy to both sides of the nucleus (Kulula \etal 1996). These jets are remarkably similar in structure to those seen in Mrk 3 and are qualitatively similar to those seen in NGC 4151. \hst ~observations of Mrk 6 have revealed, in remarkable detail, an optical emission-line jet co-spatial with the southern radio jet (Capetti \etal 1995). The similarity between Mrk 6 and NGC 4151 has been well noted in both the optical and X-ray regimes and Mrk 6 is often referred to as a ``4151 analogue''. NGC 4151 harbours one of the best studied and brightest AGN in the sky, however the complex behaviour and properties that are evident in its X-ray and optical spectra have not been observed in other nearby AGN. 

Despite its relative brightness, Mrk 6 has only been observed spectrally in the X-ray regime three times prior to the new observation presented here, once with \asca, once with \sax ~and once with \xmm. The spectral analysis of the 40 ks \asca ~observation of Mrk 6 (Feldmeier \etal 1999), despite the poor signal-to-noise of the data, revealed a 0.5-10 keV X-ray spectrum composed of a canonical underlying X-ray continuum, iron K$\alpha$ line emission, and complex, heavy column density, absorption ($\sim$10$^{23}$ cm$^{-2}$). Feldmeier \etal interpret the heavy absorption in terms of a partial covering model in which the direct line of sight passes through a relatively dust-free torus `atmosphere' with an several lines of sight, each of which pass through a significantly different column density. The X-ray spectrum was remarkably similar to the well-studied and complex Seyfert 1.5 NGC 4151, a similarity that is further heightened by the similar 0.5-10 keV X-ray luminosities of the two sources (L$_{X,2-10}$$\sim5$$\times$10$^{42}$ erg s$^{-1}$ - NGC 4151; Schurch \& Warwick 2002. {\it c.f.} L$_{X,2-10}$$\sim$1$\times$10$^{43}$ erg s$^{-1}$ - Mrk 6; Immler \etal 2003). The interpretation of Feldmeier \etal accounts for the complex absorption profile and resolves the discrepancy between the column density measured by the near infrared/optical data and the X-ray data, because the dust-free region provides photoelectric absorption but not optical extinction (Maccacaro \etal 1982). The recently published \bep ~observation yield similar details but with the backing of considerably better signal-to-noise (the LECS exposure time was $\sim$110 ks) and, thanks to the large bandpass of the \bep ~instruments, a well defined and constrained continuum (Malizia \etal 2003, Immler \etal 2003). The \bep ~data also demonstrate the presence of complex heavy X-ray absorption, which they model with a partial covering model similar to that of Feldmeier \etal (1999). Malizia \etal note, however, that the \bep ~best-fit column densities are considerably different from the column densities observed with \asca, implying that the absorption is considerably variable on the timescale of the two observations ($\sim$2 years). Malizia \etal associate the  measured X-ray column density with absorption originating in the broad line region, however it is unclear from the data whether the absorbing medium requires a partial-covering model, a warm absorption model, or indeed a combination of the two. No significant variability was detected in either the \asca ~or \sax ~observations of Mrk 6. \sax ~appears to have observed Mrk 6 in a brighter state than \asca, resulting in a somewhat different overall spectral shape. The first \xmm ~observation of Mrk 6, in conjunction with the \bep ~data, showed a similar situation, highlighting the similarities between Mrk 6 and NGC 4151 and revealing long timescale absorption variability (Immler \etal 2003). Despite having $\sim$7 times the counts of the previous \asca ~data, the \xmm ~data were still insufficient to clearly distinguish between a partial-covering model, a warm absorption model, or a combination of the two for the intrinsic absorption. The ability to distinguish between these scenarios was hampered in particular by the poor quality of the soft X-ray spectrum both in the EPIC and RGS instruments.

Here we present the most recent \xmm ~observation of Mrk 6. The observation was performed with the aim of understanding both the nature of the soft X-ray emission through the RGS data and the X-ray absorption through the EPIC instruments. This paper is organised as follows. Section~\ref{2} details the \xmm ~observation. Section~\ref{3} presents the spectra and discusses the spectral modelling of both the soft X-ray emission and the absorption. Section~\ref{4} compares the details of the current observation to the previous data and the implications that arise as a result.

\section{The \xmm ~observations}
\label{2}

Mrk 6 was observed with \xmm ~on April 26$^{th}$, 2003 (orbit 619) for a total of 55 ks. The EPIC and RGS instruments were active for the full duration of the observation however considerable prolonged background flaring was present during the first 10 ks of the observation and as a result the EPIC cameras were operated in closed mode during this period. For the remaining 45 ks of the observation the EPIC MOS and PN operated in Full Window Mode with the Medium filter in place. The data were processed with the standard EPIC and RGS processing chains incorporated in the \xmm ~SAS v6.1.0. The EPIC data were processed using the latest CCD gain values and only X-ray events corresponding to patterns 0-12 in the MOS cameras and 0-4 in the PN camera were accepted. An examination of the full field lightcurve (excluding the emission associated with Mrk 6 and other identified point sources) revealed several further periods of brief, but considerable, background flaring. Periods during which the background contribution to the count rate within a 2\arcm ~radius region centred on Mrk 6 exceeded 10\% of the source countrate in this region have been excluded from the subsequent analysis, resulting in total effective exposure times of 34 ks and 28 ks for the MOS and PN instruments respectively. A brief analysis of the impact of pile-up on the observation showed that the effect was negligible due to the moderate brightness of the source. EPIC lightcurves and spectra of Mrk 6 were extracted from a 1\arcm ~radius circle centred on the source. Background lightcurves and spectra were extracted from source free regions of the central CCD's of the MOS detectors. The background subtracted EPIC MOS lightcurve is consistent with being constant for the duration of the observation ($\chi^{2}$=445 for 410 d.o.f) with an average combined MOS count rate of $\sim$1.7 c/s. 

The RGS instruments were active for the full 55 ks duration of the observation but are subject to the same flaring periods as those observed in the EPIC cameras which, again, have been excluded from the the subsequent analysis. Dispersed RGS source and background spectra were extracted from the quiescent periods with the automated RGS extraction tasks in the SAS. The 2$^{nd}$ order RGS spectra contain no useful data, primarily because of the relatively faint nature of the source, and as a result only the 1$^{st}$ order spectra are presented here. The resulting RGS spectra have a total effective exposure time of 34 ks for both the RGS 1 and 2 instruments. 

Spectral fitting was carried out using the XSPEC v11.3 software package with instrument response matrices generated by the SAS software for each of the spectra independently. The source spectra were grouped to a minimum of 20 counts per bin to allow the use of $\chi$$^{2}$ minimisation techniques.

\begin{figure}
\centering
\begin{minipage}{85 mm}
\centering
\vbox{
\includegraphics[height=8 cm, angle=270]{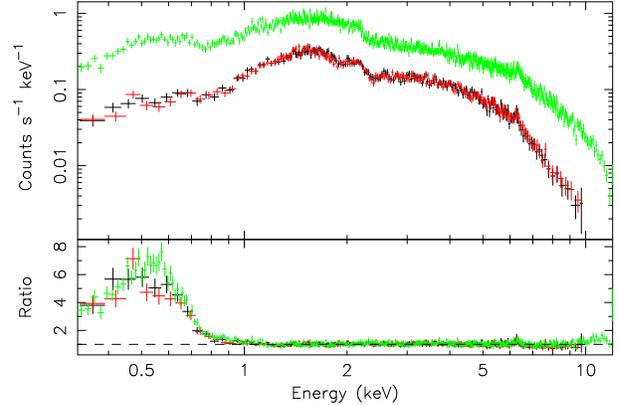}
}
\caption{The EPIC MOS (Black - MOS 1, Red - MOS 2) and PN (Green) spectra of Mrk 6. The lower panel shows the ratio of the data to a basic absorbed power-law fit.}
\label{EPICfig1}
\end{minipage}
\end{figure}

\section{Modelling the \xmm ~spectra}
\label{3}

\subsection{The EPIC Spectra}
\label{3-1}

We begin by fitting the EPIC MOS and PN data between 1-5 keV and 7.5-10 keV with a basic absorbed power-law model. This model, while obviously unrealistic, allows us to characterise the basic properties of the X-ray spectrum, particularly for comparison with other AGN. Figure~\ref{EPICfig1} shows the EPIC MOS and PN spectra of Mrk 6, along with the ratio of the data to a basic absorbed power-law fit.  The fit is characterised by a very flat ($\Gamma$$\sim$1.2) continuum and absorption an order of magnitude above the level of Galactic absorption in the direction of Mrk 6 (7$\times$10$^{21}$ cm$^{-2}$ c.f. 6.4$\times$10$^{20}$ cm$^{-2}$ - Stark \etal 1992). The positive residuals around 6.4 keV are indicative of the presence of a neutral iron K$\alpha$ line and the large positive residuals below 0.8 keV demonstrate the presence of a strong `soft X-ray excess', a feature that is common to many AGN ({\it e.g.} Piro, Matt \& Ricci - 1997, Gierli{\' n}ski \& Done - 2004). The extremely flat continuum in this simple fit is suggestive of a strong Compton reflection component, coupled with a steeper underlying continuum. Indeed, analysis of the high energy ($>$10 keV) X-ray spectrum of Mrk 6 from a previous \bep ~observation confirms this, detecting a steeper underlying continuum (1.6$<$$\Gamma$$<$2.0) and a Compton reflection component with a reflection fraction of $\sim$1.0 (Malizia \etal 2003; Immler \etal 2003). 

Both the MOS and PN spectra show similar large-scale features with one exception highlighted by the fit to the absorbed power-law model. Between 0.5-0.6 keV, the PN shows a considerable difference in spectral shape to both MOS spectra which both show a dip in their spectra. An investigation of the RGS 1 spectra supports the MOS spectral shape in this region (the RGS 2 spectrum suffers from a defective CCD across this energy range). We initially limit our modelling to the 0.8-12 keV region for both the MOS and PN cameras to exclude the soft X-ray excess and simplify the analysis of the continuum, given the discrepancy between the PN and the other instruments we retain the 0.8 keV lower limit for the PN data in the broadband spectral fits detailed in Section~\ref{3-3}.

We define a base model, based on the phenomenological properties observed in the over-simplistic absorbed power-law fit, with the following components:  

\begin{itemize}

\item  A power-law continuum. The continuum slope is initially fixed to be $\Gamma$=1.81, as determined from the analysis of the high energy \bep ~data by Immler \etal (2003), while the normalisation, A$_{1}$, remains a free parameter in the fits.

\item A neutral Compton reflection component (modelled by PEXRAV in XSPEC, Magdziarz \& Zdziarski 1995). The reflection scaling factor, $R$, is fixed to be 1.21, the best-fit value found from the analysis of the \bep ~data by Immler \etal (2003) as are the other remaining parameters ({\it i.e.} incident continuum slope, $\Gamma$=1.81; Normalisation, A$_{2}$=4.68$\times$10$^{-3}$ $\xspnorm$; Inclination cos $i$=0.94). In addition the metal abundance in the reflector was fixed at the solar value.

\item Complex absorption. Two distinct models were used for the nature and form of the complex absorption. The first model is the partial covering model favoured in previous analyses of the X-ray spectrum of Mrk 6. The second model is a layered model that incorporates a heavy column density ({\it e.g.} $\sim$10$^{23}$ cm$^{-2}$) partially ionised absorber, a model that has been extremely successful in modelling the detailed X-ray spectrum of NGC 4151. The details of the absorption modelling are the primary focus of the following discussion and are given in detail there.

\item  An intrinsically narrow iron K$\alpha$ emission line of intensity $I_{K\alpha}$, at an energy $E_{K\alpha}$.

\item  Galactic absorption (6.4$\times$10$^{20}$ cm$^{-2}$). 

\end{itemize}

\begin{figure}
\centering
\begin{minipage}{85 mm}
\centering
\vbox{
\includegraphics[height=8 cm, angle=270]{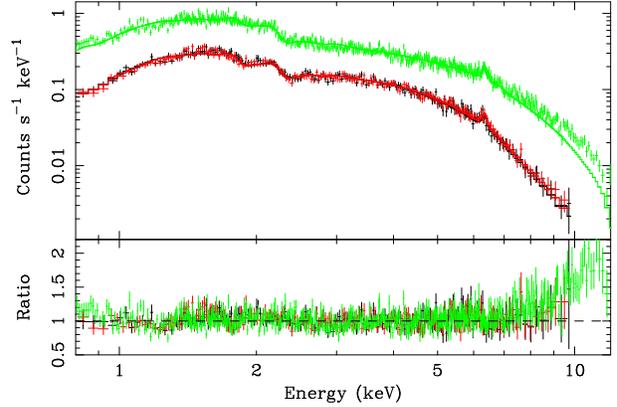}
}
\caption{The EPIC spectra of Mrk 6 fit with the base model incorporating partial covering absorption.}
\label{EPICfig2}
\end{minipage}
\end{figure}

\begin{table*}
%\centering
\begin{minipage}{170 mm} 
\centering
\caption{The best-fitting EPIC spectral parameters}
\begin{tabular}{lccccc}
Parameter & PC1 & PC2 & PC3 & WA1 & WA2\\ \hline

N$_{H,1}$$^{a}$ & 0.57$^{+0.03}_{-0.03}$ & 0.44$^{+0.05}_{-0.05}$ & 0.28$^{+0.04}_{-0.03}$ & 0.47$^{+0.02}_{-0.03}$ & 0.028$^{b}$\\ 
$f_{cov, 1}$ & 0.43$^{+0.05}_{-0.04}$ & 0.37$^{+0.05}_{-0.05}$ & 0.36$^{+0.03}_{-0.04}$ & - & - \\
 & & & & & \\
N$_{H,2}$$^{a}$ & 3.95$^{+0.35}_{-0.31}$ & 2.22$^{+0.26}_{-0.23}$ & 1.39$^{+0.06}_{-0.11}$ & 7.48$^{+0.25}_{-0.19}$ & 3.16$^{+0.06}_{-0.6}$ \\
log$\xi$ & - & - & - & 2.455$^{+0.007}_{-0.006}$ & 1.897$^{+0.039}_{-0.031}$\\ 
$f_{cov, 2}$ & 0.57$^{+0.04}_{-0.05}$ & 0.63$^{+0.05}_{-0.05}$ & 0.64$^{+0.04}_{-0.03}$ & - & - \\
 & & & & & \\
$\Gamma$$_{pl}$ & 1.81$^{c}$ & 1.81$^{c}$ & 1.44$^{+0.09}_{-0.07}$ & 1.81$^{c}$ & 1.81$^{c}$\\
A$_{pl}$$^{d}$ & 5.91$^{+0.26}_{-0.22}$ & 4.74$^{+0.42}_{-0.44}$ & 3.10$^{+0.20}_{-0.21}$ & 6.18$^{+0.07}_{-0.07}$ & 4.67$^{+0.10}_{-0.08}$ \\
R & 1.21$^{c}$ & 4.00$^{+0.32}_{-0.32}$ & 0.60$^{+0.14}_{-0.14}$ & 1.21$^{c}$ & 4.3$^{+0.2}_{-0.3}$ \\
E$_{K\alpha}$ & 6.44$^{+0.03}_{-0.02}$ & 6.44$^{+0.03}_{-0.03}$ & 6.44$^{+0.03}_{-0.02}$ & 6.45$^{+0.02}_{-0.03}$ & 6.44$^{+0.04}_{-0.03}$ \\ 
I$_{K\alpha}$$^{e}$ & 1.8$^{+0.3}_{-0.4}$ & 1.0$^{+0.4}_{-0.3}$ & 1.34$^{+0.36}_{-0.34}$ & 1.62$^{+0.33}_{-0.34}$ & 0.93$^{+0.37}_{-0.34}$ \\
 & & & & \\
$\chi^{2}$ & 2337 & 2161 & 2117 & 2365 & 2172\\
d.o.f & 2263 & 2262 & 2261 & 2264 & 2263 \\ \hline
\multicolumn{2}{l}{\scriptsize $^{a}$ $\times$10$^{22}$ cm$^{-2}$} & \multicolumn{2}{l}{\scriptsize $^{d}$ 10$^{-3}$ photon keV$^{-1}$ cm$^{-2}$ s$^{-1}$} \\
\multicolumn{2}{l}{\scriptsize $^{b}$ Upper Limit.} & \multicolumn{2}{l}{\scriptsize $^{e}$ 10$^{-5}$ photon cm$^{-2}$ s$^{-1}$} \\
\multicolumn{2}{l}{\scriptsize $^{c}$ Fixed parameter.} & \\
\end{tabular}
\hspace{-1 cm}
\label{EPICfit}
\end{minipage}
\end{table*}

\subsubsection{The partial covering model}
\label{3-1-1}

\begin{figure}
\centering
\begin{minipage}{85 mm}
\centering
\vbox{
\includegraphics[height=8 cm, angle=270]{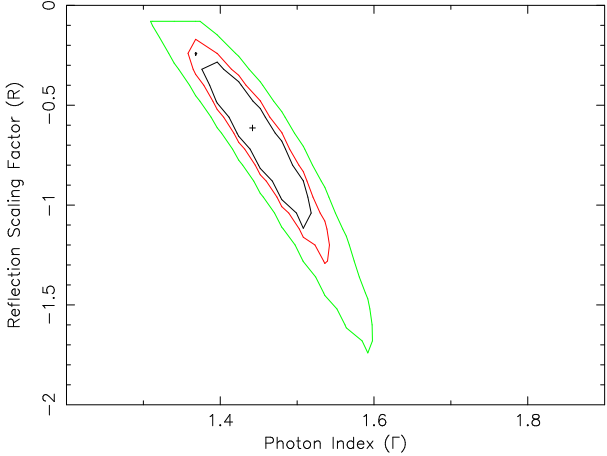}
}
\caption{Confidence contour plot of $\Gamma$ and $R$ for Model PC3. The black, red \& green lines are confidence contours corresponding to 1, 2 and 3$\sigma$, respectively ($\Delta$$\chi$$^{2}$=2.71, 4.0 \& 9.0). The diagonal nature of the contours highlights the strong degeneracy between the parameters. }
\label{contplot}
\end{minipage}
\end{figure}

As discussed in Section~\ref{1}, previous approaches to modelling the absorption intrinsic to Mrk 6 have been largely based on a partial covering model ({\it e.g.} Feldmeier \etal 1999; Immler \etal 2003). In partial covering models, a fraction, $f_{cov}$, of the hard power-law continuum is absorbed by a cold gas with column density $N_{H,1}$, whilst the remaining fraction of the continuum, $1-f_{cov}$, is absorbed by a considerably different column density $N_{H,2}$. This absorption profile was incorporated into the base model defined previously and fit to the three EPIC datasets simultaneously, resulting in a reasonable fit ($\chi$$^{2}$=2337 for 2263 d.o.f; Figure~\ref{EPICfig2}; Table~\ref{EPICfit} - Model PC1). Despite the fit being acceptable, there is a clear deviation of the model from the data above energies of $\sim$8 keV. The deviation is particularly evident in the PN data, but is also observed to a lesser degree in the MOS datasets. The high energy deviation is suggestive of a mismodelling of the Compton reflection continuum which, given the tight constraints employed in the base model, is not surprising. Freeing up the relative reflection component, R, results in a considerable improvement to the fit ($\Delta\chi^{2}$=176 for 1 d.o.f; Table~\ref{EPICfit} - Model PC2) with a reflection component $\sim$3.3 times stronger than the previous measurement.  Despite the considerable improvement of the fit the PN data $>$9 keV still lie systematically above the model. It is not clear that the PN calibration $>$10 keV is well understood, however under the assumption that it is, the under-prediction of the model thus suggests that the continuum is incorrectly modelled by the tightly constrained power-law model. A careful investigation of $\chi^{2}$-space when the photon index of the underlying continuum, $\Gamma$, is allowed to be a free component in the fit reveals a considerably better fit to the data ($\Delta\chi^{2}$=44 for 1 d.o.f; Table~\ref{EPICfit} - Model PC3) with a resulting best-fit power-law photon index of $\sim$1.44 and a considerably weaker contribution from Compton reflection (R$\sim$0.6). The surprisingly flat slope appears to be well constrained, however the corresponding change in the strength of the inherently flat Compton reflection indicates a strong degeneracy between the two parameters shown clearly by Figure~\ref{contplot}.

We reiterate that the discrepancy between the model and the data was only present in the PN data for the previous fits and it is possible that a calibration difference is responsible for the over-prediction. Allowing the relative reflection component and the continuum photon index to take different values for the fit to the PN data than for the MOS data to test this possibility resulted in two considerably different best-fit values for the continuum photon index and Compton reflection. The best-fit MOS values are $\Gamma$=1.40$^{+0.10}_{-0.02}$, R$<$0.27, in contrast to the best-fit PN values of $\Gamma$=1.57$^{+0.09}_{-0.02}$, R=1.0$^{+0.5}_{-0.2}$, again highlighting the degeneracy between photon index and Compton reflection. 

We stress that the fits with $\Gamma$$\sim$1.4 and $\Gamma$$>$1.6 both provide a statistically acceptable fit to the data and, given that such a flat continuum slope is incompatible with the previous wide bandpass \bep ~data (which in several analyses has measured a best-fit photon index of $\Gamma$$>$1.6 using similar models to that presented here), we argue that Model PC2, despite a high contribution from Compton reflection and residuals in the PN fit $>$9 keV, is a more plausible physical representation of the data.

\begin{figure}
\centering
\begin{minipage}{85 mm}
\centering
\vbox{
\includegraphics[height=8 cm, angle=270]{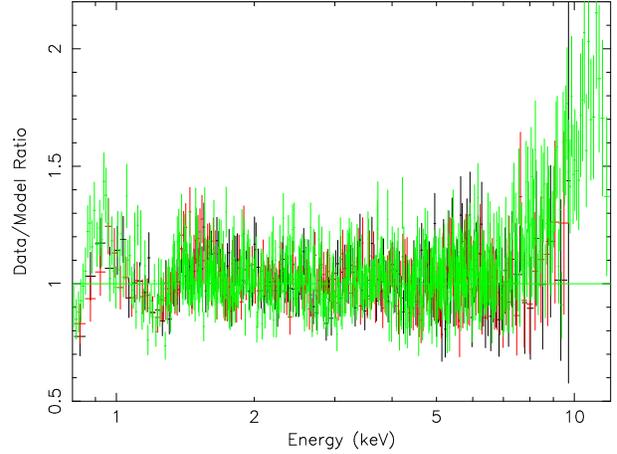}
}
\caption{The Data/Model ratio residuals resulting from fitting the EPIC spectra of Mrk 6 with the base model incorporating partially ionised absorption.}
\label{EPICfig3}
\end{minipage}
\end{figure}

\subsubsection{The layered model}
\label{3-1-2}

The second approach to modelling the absorption intrinsic to Mrk 6 utilises a model in which a heavy column density partially ionised absorber is layered with a lower column density neutral absorbing column. This model has been extremely successful in detailing the X-ray spectrum of NGC 4151 (Schurch \& Warwick 2002, Schurch \etal 2003). In this model the absorption affects all the power-law continuum flux, instead introducing the required complexity in the absorption profile through the details of the ionised absorption. Here, we use the same warm absorption model, calculated by XSTAR, that was used in the Schurch \etal (2003) analysis of NGC 4151. This absorption profile was incorporated into the base model defined previously and fit to the three EPIC datasets simultaneously, resulting in an acceptable, if marginally worse, fit to that of the initial tightly constrained partial covering model ($\chi$$^{2}$=2365 for 2264 d.o.f; Figure~\ref{EPICfig3}; Table~\ref{EPICfit} - Model WA1). Clearly the fit suffers from the same deviations from the data at high energies that were observed with the partial covering model fit, along with somewhat more pronounced residuals $<$1.5 keV. The change in the model of the absorption profile does not strongly impact on the model fit $>$5 keV, as highlighted by the reaction of this fit when the relative reflection parameter is allowed to be a free parameter (Table~\ref{EPICfit} - Model WA2); the fit improves considerably ($\Delta\chi^{2}$=89 for 1 d.o.f) and the reflection component is $\sim$3.6 times stronger, a similar effect to that previously observed with the partial covering model. The increase in the strength of the reflection component is coupled with considerable decreases in both the ionisation state of the warm absorber and its column density. Similar effects to those detailed in the previous section are also observed when the photon index, $\Gamma$, is available as a free parameter in the fit and when the parameter values for the PN data are decoupled from the values for the MOS data.

\subsection{The RGS spectra}
\label{3-2}

The only appreciable difference between the best-fits of the two absorption models to the EPIC data is the presence of more significant residuals in the 0.8-1.5 keV band in the warm absorber best-fit. Given the apparent similarity between Mrk 6 and NGC 4151 it is not unreasonable to suspect that the soft X-ray emission in Mrk 6 is composed almost entirely of discrete line emission similar to that in NGC 4151 (Schurch \etal 2004, Ogle \etal 2000). In such a scenario, the presence of increased residuals around 0.8-1.5 keV can be directly attributed to the presence of this line emission (specifically, Magnesium and Neon emission lines). To investigate this possibility in detail it is prudent to investigate the RGS data, with a view to identifying specific emission line features.

\begin{figure}
\centering
\begin{minipage}{85 mm}
\centering
\vbox{
\includegraphics[height=8 cm, angle=270]{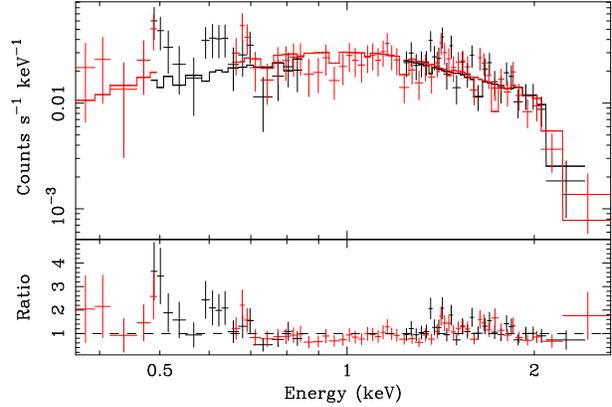}
}
\caption{The RGS (Black - RGS 1, Red - RGS 2) spectra of Mrk 6. The lower panel shows the ratio of the data to the basic absorbed power-law fit in the upper panel. The gaps in the data correspond to the energy ranges covered by the failed CCDs in the RGS detectors (RGS1: $\sim$0.8-1.25 keV, RGS2: $\sim$0.5-0.66 keV)}.
\label{RGSfig1}
\end{minipage}
\end{figure}

The RGS data are shown in Figure~\ref{RGSfig1} along with the ratio of the data to a basic absorbed power-law fit, in a similar fashion to Figure~\ref{EPICfig1}. The RGS spectra clearly demonstrate that the soft X-ray emission in Mrk 6 is dominated by some form of continuum emission, very different in nature to the soft X-ray line emission in NGC 4151. Despite this difference, the emission does show features consistent with possible line emission or edge absorption, particularly in the 0.4-0.7 keV and 1.3-1.9 keV regions. modelling of the RGS data independently of the EPIC data reveals that the soft X-ray spectrum is not well fit with simple models including a thermal plasma ($\chi$$^{2}$=280 for 95 d.o.f), emission from a blackbody ($\chi$$^{2}$=115 for 95 d.o.f) or a simple absorbed power-law ($\chi$$^{2}$=179 for 96 d.o.f). Following these simple fits we fit the RGS spectra with the most reasonable fits from the EPIC analysis (Models PC2 and WA2). Most of the free parameters in the RGS fits were fixed to the values from the best-fits to the EPIC data, either because the components of the model they refer to are outside the energy range of the RGS ({\it e.g.} the iron K$\alpha$ line) or the components are broadband continuum components that are considerably better constrained by the EPIC data. In the best-fit partial covering model (PC2), the power-law normalisations and the lowest column density absorber were allowed to be free parameters in the fit. The model provides a barely reasonable fit to the RGS data ($\chi^{2}$=130 for 95 d.o.f); in addition, the column density of the absorption drops to $\sim$2$\times$10$^{21}$ cm$^{-2}$, well below the best-fit value from the EPIC data, and the model leaves considerable residuals around the relatively complex 0.4-0.7 keV and 1.3-1.9 keV regions. Adding a thermal plasma component (MEKAL) does not significantly improve the fit or allow the low column density to take on a more reasonable value. Allowing the chemical abundance of the thermal plasma to be a free parameter in the fit does improve the fit but results in a chemical abundance $\leq$3\% of solar abundance values, essentially removing all of the line emission from the model. Adding a low temperature blackbody component in place of the thermal plasma provides a considerable increase in the quality of the fit (Table~\ref{RGStab}) and may represent the tail of the termal X-ray emission from a the accretion disk. We note that the properties of the black body component in the warm absorber model are poorly constrained due to an obvious degeneracy between the normalisation of the power-law continuum and the temperature of the black body emission.

By contrast the best-fit EPIC warm absorption model (WA2) provides an excellent representation of the RGS data ($\chi^{2}$=83 for 93 d.o.f) without the addition of new spectral components. In this fit both the warm and cold column densities, along with the power-law normalisation, remain free parameters. The fit differs from the best-fit EPIC values, suggesting a slightly more ionised medium, a smaller column density of warm material and considerably more cold material long the line of sight (Table~\ref{RGStab}). The considerably larger cold column density is required to prevent the warm absorber model over-predicting the soft X-ray emission.

\begin{table}
%\centering
\begin{minipage}{85 mm} 
\centering
\caption{The best-fitting RGS spectral parameters}
\begin{tabular}{lcc}
Parameter & PC2+Blackbody & WA2 \\ \hline
N$_{H,cold}$$^{a}$ & 1.04$^{+0.22}_{-0.46}$ & 0.22$^{+0.05}_{-0.11}$ \\
N$_{H,warm}$$^{a}$ & - & 2.3$^{+1.1}_{-1.1}$ \\
log$\xi$ & - & 1.93$^{+0.17}_{-0.26}$ \\ 
A$_{pl}$$^{c}$ & 3.6$^{+0.7}_{-2.1}$ & 4.63$^{+1.06}_{-0.78}$ \\
kT$^{d}$ & 0.15$^{+0.04}_{-0.03}$ & - \\
A$_{bb}$$^{e}$ & 0.92$^{+0.16}_{-0.17}$ & - \\
& & \\
$\chi^{2}$ & 76 & 83 \\
d.o.f & 93 & 93 \\ \hline
\multicolumn{2}{l}{\scriptsize $^{a}$ $\times$10$^{22}$ cm$^{-2}$} \\
\multicolumn{2}{l}{\scriptsize $^{b}$ Upper Limit}\\
\multicolumn{2}{l}{\scriptsize $^{c}$ 10$^{-3}$ photon keV$^{-1}$ cm$^{-2}$ s$^{-1}$} \\
\multicolumn{2}{l}{\scriptsize $^{d}$ keV$^{-1}$} \\
\multicolumn{2}{l}{\scriptsize $^{e}$ 10$^{-5}$ photon keV$^{-1}$ cm$^{-2}$ s$^{-1}$} \\
\end{tabular}
\hspace{-1 cm}
\label{RGStab}
\end{minipage}
\end{table}

\subsection{Joint, broadband, spectral fitting}
\label{3-3}

\begin{figure}
\centering
\begin{minipage}{85 mm}
\centering
\vbox{
\includegraphics[height=8 cm, angle=270]{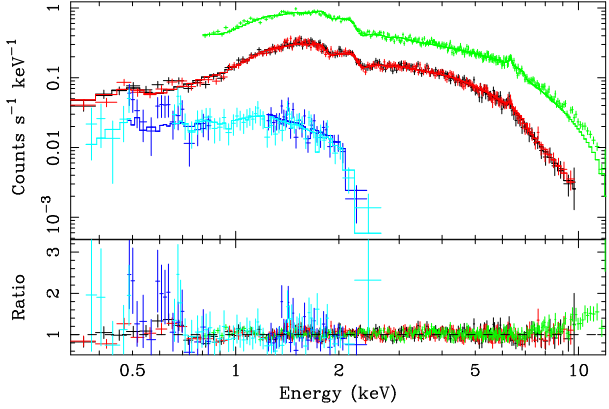}
}
\caption{The EPIC and RGS spectra of Mrk 6 fit with the best-fit partial covering absorption model including emission from a blackbody.}
\label{jointfig1}
\end{minipage}
\end{figure}

\begin{figure}
\centering
\begin{minipage}{85 mm}
\centering
\vbox{
\includegraphics[height=8 cm, angle=270]{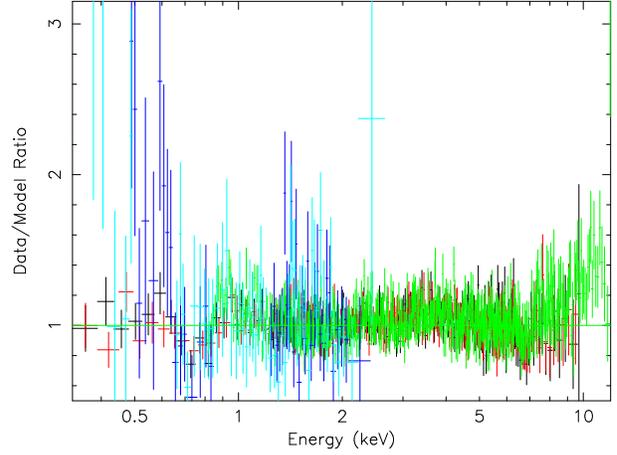}
}
\caption{The Data/Model ratio residuals resulting from fitting the EPIC and RGS spectra of Mrk 6 with the best-fit warm absorption model.}
\label{jointfig2}
\end{minipage}
\end{figure}

The RGS spectra for Mrk 6 highlight the importance of fitting the full energy range of the spectra when using warm absorption models unless there is unambiguous evidence that the leaked flux from such a component is not present in the soft X-ray emission (as is the case in NGC 4151). Taking into account the details of the best-fits to the RGS data, we performed joint spectral fits on the EPIC and RGS data over the full 0.3-12 keV energy range for the MOS detectors and a more limited energy range for the PN detector (the PN data up to 0.8 keV remained excluded for the reasons discussed in Section~\ref{3-1}). The initial models fit to the joint EPIC and RGS spectra have the same free parameters listed earlier for the WA2 and PC2 models however, in keeping with the RGS fits, a simple blackbody model was added to the PC2 model. Figures~\ref{jointfig1} \&~\ref{jointfig2} show the best-fits of the partial covering absorption model and the warm absorber model to the EPIC and RGS data, respectively. Both models are a good statistical representation of the data with reasonable physical parameters ($\chi^{2}$=2327 and 2360 for 2419 and 2422 d.o.f, respectively). Details of the fits are given in Table~\ref{jointfits}. The warm absorber fit is somewhat improved by the addition of a scattered power-law component ($\Delta\chi^{2}$=37 for 1 d.o.f) however the statistically good nature of the fit prior to the addition of the scattered power-law prevents us from concluding that this component is {\it required} in the modelling. When included, the scattered power-law contains $\sim$3.5\% of the flux of the underlying continuum.

\begin{table}
%\centering
\begin{minipage}{85 mm} 
\centering
\caption{The best-fit parameters for the joint EPIC and RGS spectra}
\begin{tabular}{lcc}
Parameter & PC2+bb & WA2\\ \hline

N$_{H,1}$$^{a}$ & 0.89$^{+0.12}_{-0.13}$ & 0.27$^{+0.1}_{-0.01}$ \\
$f_{cov, 1}$ & 0.69$^{+0.08}_{-0.10}$ & - \\
 & & \\
N$_{H,2}$$^{a}$ & 3.8$^{+1.3}_{-0.9}$ & 2.89$^{+0.14}_{-0.14}$ \\
log$\xi$ & - & 2.069$^{+0.025}_{-0.021}$ \\
$f_{cov, 2}$ & 0.31$^{+0.10}_{-0.08}$ & - \\
 & & \\
$\Gamma$$_{pl}$ & 1.81$^{c}$ & 1.81$^{c}$ \\
A$_{pl}$$^{d}$ & 4.48$^{+0.78}_{-0.91}$ & 4.6$^{+0.1}_{-0.1}$ \\
R & 3.47$^{+0.34}_{-0.38}$ & 4.44$^{+0.26}_{-0.26}$ \\
E$_{K\alpha}$ & 6.44$^{+0.03}_{-0.03}$ & 6.44$^{+0.03}_{-0.03}$ \\
I$_{K\alpha}$$^{e}$ & 0.98$^{+0.33}_{-0.38}$ & 0.91$^{+0.35}_{-0.34}$ \\
 & & \\
kT$^{d}$ & 0.17$^{+0.03}_{-0.02}$ & - \\
A$_{bb}$$^{f}$ & 0.68$^{+0.16}_{-0.17}$ & - \\
& & \\
$\chi^{2}$ & 2327 & 2360 \\
d.o.f & 2419 & 2422 \\ \hline
\multicolumn{2}{l}{\scriptsize $^{a}$ $\times$10$^{22}$ cm$^{-2}$} \\
\multicolumn{2}{l}{\scriptsize $^{b}$ Upper Limit.} \\
\multicolumn{2}{l}{\scriptsize $^{c}$ Fixed parameter.} \\
\multicolumn{2}{l}{\scriptsize $^{d}$ 10$^{-3}$ photon keV$^{-1}$ cm$^{-2}$ s$^{-1}$} \\
\multicolumn{2}{l}{\scriptsize $^{e}$ 10$^{-5}$ photon cm$^{-2}$ s$^{-1}$} \\
\multicolumn{2}{l}{\scriptsize $^{f}$ 10$^{-5}$ photon keV$^{-1}$ cm$^{-2}$ s$^{-1}$} \\
\end{tabular}
\hspace{-1 cm}
\label{jointfits}
\end{minipage}
\end{table}

\section{Discussion}
\label{4}

The analysis presented in Section~\ref{3} demonstrates that the absorption profile of Mrk 6 is extremely complex and that distinguishing between the plausible absorption profiles is a tricky endeavour. Despite the partial coverer providing a marginally better broadband fit to the \xmm ~data, both the partial covering model and the warm absorption model are good representations of the data with reasonable physical parameters and there is little to distinguish between the two.

\subsection{A comparison with previous \xmm ~data (1997)}
\label{4-1}

Any model of the absorption in Mrk 6 needs to be both a good statistical representation of the data in any specific observation and needs to be able to explain any observed absorption variability in a physically plausible manner. We stress the latter clause because a comparison of the results from the spectral fitting from the observation presented here to the previous \xmm ~observation clearly indicates that considerable N$_{H}$ changes are evident in Mrk 6, regardless of which absorption model you use to represent the data. 

\subsubsection{Partial covering}
\label{4-1-1}

Analysis of the previous X-ray spectra from Mrk 6 have favoured a `duel partial covering' model, the details of which are considerably more complex than the partial covering model presented here (Feldmeier \etal 1999, Immler \etal 2003 \& Malizia \etal 2003). The model used in Section~\ref{3-1} defines two distinct lines of sight to the source, both with different intrinsic amounts of absorption. Such a model is only plausible in the situation where the absorbing medium is sufficiently close to the continuum source that the continuum source does not appear point-like with respect to the absorbing material and that the region over which the column density changes significantly is small ({\it i.e.} that the edges of absorbing clouds are relatively sharp). Taking the assumption that the X-ray continuum emission originates in the inner regions of the accretion disk, this would imply that the absorption originates in the broad-line region clouds, or closer material (for a 10$^{8}$\Msun black hole, an emission region of 5R$_{s}$ corresponds to $\sim$3\arcm at a distance of 0.1 pc, compared to 20\arcs at a distance of 1pc). 

In fact, the `duel partial covering' models favoured in previous modelling of Mrk 6 define a total of four unique lines of sight, three with different, distinct, intrinsic absorption and one intrinsically unabsorbed line of sight. Despite the good statistical fit found with a considerably less complex model, involving only two distinct lines of sight, we fitted the \xmm ~spectra with the more complex model for the purposes of direct comparisons with the previous studies. This model (Table~\ref{2PCfit} - \xmm ~2003) results in fit of similar quality to the initial partial covering model ($\chi$$^{2}$=2330 for 2363 d.o.f). Examining the details of the four different lines of sight in this model, it becomes clear that two of the absorbed lines of sight are strongly dominant, accounting for a total of 50\% and 47\% of the continuum flux and with column densities of $\sim$4.5$\times$10$^{22}$ and $\sim$0.63$\times$10$^{22}$ cm$^{-2}$, respectively. These values are extremely similar to the values for the two absorbed lines of sight used in the initial partial covering model ({\it c.f.} Table~\ref{EPICfit} - Model PC1) confirming that despite greater complexity the `double partial covering' model closely resembles the simpler fit. Similarly, adding an unabsorbed continuum component to the simple initial partial covering model does not improve the spectral fit considerably ($\Delta\chi^{2}$=1 for 1 d.o.f) and results only in an upper limit of 3.5$\times$10$^{-5}$$\xspnorm$ on the flux in the unabsorbed component ($<$0.3\% of the flux in the absorbed components), again confirming that there is little additional complexity in the X-ray spectrum $>$0.8 keV. 

\begin{table}
%\centering
\begin{minipage}{85 mm} 
\centering
\caption{The `dual partial covering' model}
\begin{tabular}{lcc}
Parameter & \xmm ~2001 & \xmm ~2003 \\ \hline

N$_{H,1}$$^{a}$ & 10.32$^{+0.72}_{-0.56}$ & 4.43$^{+0.17}_{-0.33}$ \\ 
$f_{cov, 1}$ & 0.51$^{+0.02}_{-0.02}$ & 0.50$^{+0.01}_{-0.01}$ \\
N$_{H,2}$$^{a}$ & 8.05$^{+0.69}_{-0.48}$ & 3.8$^{+0.18}_{-0.35}$ \\
$f_{cov, 2}$ & 0.05$^{+0.01}_{-0.01}$ & 0.02$^{+0.01}_{-0.01}$ \\
N$_{H,3}$$^{a}$ & 2.27$^{+0.03}_{-0.08}$ & 0.63$^{+0.01}_{-0.01}$ \\ 
$f_{cov, 3}$ & 0.40$^{+0.01}_{-0.02}$ & 0.46$^{+0.01}_{-0.01}$ \\
N$_{H,4}$$^{b}$ & 0 & 0\\
$f_{cov}$ & 0.04$^{+0.01}_{-0.01}$ & 0.02$^{+0.01}_{-0.01}$ \\
 & & \\
$\Gamma$$_{pl}$ & 1.81$^{c}$ & 1.81$^{c}$\\
A$_{pl}$$^{d}$ & 5.39$^{+0.05}_{-0.06}$ & 5.98$^{+0.04}_{-0.03}$ \\
R & 1.21$^{+0.14}_{-0.16}$ & 1.21$^{c}$ \\
E$_{K\alpha}$ & 6.47$^{+0.03}_{-0.03}$ & 6.45$^{+0.02}_{-0.03}$ \\ 
I$_{K\alpha}$$^{e}$ & 1.68$^{+0.39}_{-0.53}$ & 1.7$^{+0.3}_{-0.4}$ \\
 & & \\
$\chi^{2}$ & 813 & 2330 \\
d.o.f & 885 & 2263 \\ \hline
\multicolumn{2}{l}{\scriptsize $^{a}$ $\times$10$^{22}$ cm$^{-2}$} \\
\multicolumn{2}{l}{\scriptsize $^{b}$ Unabsorbed line of sight} \\
\multicolumn{2}{l}{\scriptsize $^{c}$ Fixed parameter.} \\
\multicolumn{2}{l}{\scriptsize $^{d}$ 10$^{-3}$ photon keV$^{-1}$ cm$^{-2}$ s$^{-1}$} \\
\multicolumn{2}{l}{\scriptsize $^{e}$ 10$^{-5}$ photon cm$^{-2}$ s$^{-1}$} \\
\end{tabular}
\hspace{-1 cm}
\label{2PCfit}
\end{minipage}
\end{table}

A direct comparison between the best-fit column density values for the 2001 and 2003 observations (Table~\ref{2PCfit}) reveals a factor of $\sim$2-3.5 drop in the column density during the intervening period. Remarkably, despite such a large column density variation, the partial covering fits still require several lines of sight to the source and, furthermore, the best-fit lines of sight are very similar to those present in the earlier observation. This argues strongly against the partial covering model being the correct interpretation of the complex absorption since it is difficult to conceive of a situation in which absorption in clouds close to the source results in large column density variations without considerable differences in the covering fractions of the clouds.

\subsubsection{Warm absorption}
\label{4-1-2}

Despite not being the favoured model for previous observations of Mrk 6, the warm absorption model does provide a good fit to the \xmm ~data. Comparing the details of the current warm absorber model parameters to those given in Immler \etal (2003) for the 2001 observation reveals considerable changes in both ionisation and column density of the absorption (log$\xi_{'01}$=2.42, log$\xi_{'03}$=2.07; N$_{H, '01}$=1.33$\times$10$^{23}$ cm$^{-2}$, N$_{H, '03}$=0.29$\times$10$^{23}$ cm$^{-2}$). Assuming that the absorption originates in a single `cloud' ({\it i.e.} a region of material with a single ionisation state and density), these differences imply very different properties for the cloud seen in the 2001 observation than for the cloud present during the 2003 observation. In particular, the considerable change in ionisation state implies either a large change in the radial distance, or a large change in the cloud density.

An alternative interpretation can be made in terms of two separate ionised regions, a uniform low column density, low ionisation region and an irregular higher column density higher ionisation region. There is considerable evidence for absorption with multiple ionisation states in Seyferts, although as yet no multiple warm absorbers have been identified with such a heavy column density. In this model, the lower column density, lower ionisation state region remains constant between both observations, while the second region (comprising a higher column density, higher ionisation state) is only present along the line of sight during the first \xmm ~observation and has a higher ionisation state either as a result of a closer proximity to the ionising continuum emission or as a result of a lower density. We note that this is {\it not} the same as the partial covering model since it requires only one line of sight.

\begin{figure}
\centering
\begin{minipage}{85 mm}
\centering
\vbox{
\includegraphics[height=8 cm, angle=270]{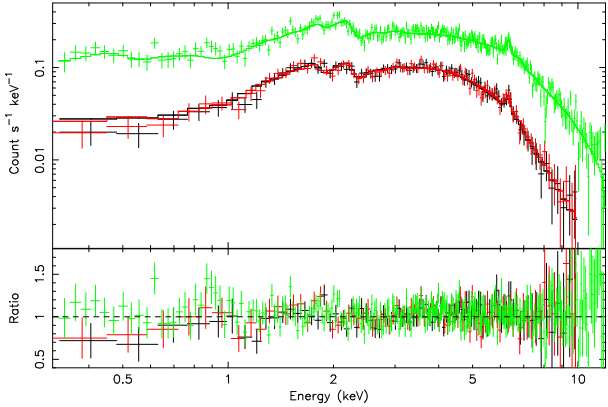}
}
\caption{The 2001 EPIC spectra of Mrk 6 fit with the best-fit multi-zone warm absorber model.}
\label{mzWA1}
\end{minipage}
\end{figure}

Fitting the EPIC PN and MOS spectra from the 1997 observations with this multi-zone warm absorber model (Figure~\ref{mzWA1} \& Table~\ref{mzWAfit}) results in an excellent fit to these data with reasonable physical parameters and, with the heavier warm absorbing material no longer present along the line of sight, also results in a good fit to the current \xmm ~data. 

\begin{table}
%\centering
\begin{minipage}{85 mm} 
\centering
\caption{The best-fit parameters for the \xmm ~2001 spectra}
\begin{tabular}{lcc}
Parameter & Multi-zone warm absorber\\ \hline

N$_{H,1}$$^{a}$ & 11.6$^{+1.4}_{-2.1}$ \\
log$\xi$$_{1}$ & 2.512$^{+0.031}_{-0.052}$ \\
 & & \\
N$_{H,2}$$^{b}$ & 2.89 \\
log$\xi$$_{2}$$^{b}$ & 2.069 \\
 & & \\
N$_{H,cold}$$^{a}$ & 1.25$^{+0.25}_{-0.35}$ \\
$\Gamma$$_{pl}$ & 1.81$^{b}$ \\
A$_{pl}$$^{c}$ & 4.39$^{+0.16}_{-0.34}$ \\
R & 1.93$^{+0.52}_{-0.59}$ \\
A$_{scat pl}$$^{d}$ & 1.0$^{+0.07}_{-0.18}$ \\
E$_{K\alpha}$$^{b}$ & 6.44 \\
I$_{K\alpha}$$^{e}$ & 1.47$^{+0.38}_{-0.46}$ \\
 & & \\
$\chi^{2}$ & 1866 \\
d.o.f & 2425 \\ \hline
\multicolumn{2}{l}{\scriptsize $^{a}$ $\times$10$^{22}$ cm$^{-2}$} \\
\multicolumn{2}{l}{\scriptsize $^{b}$ Fixed parameter.} \\
\multicolumn{2}{l}{\scriptsize $^{c}$ 10$^{-3}$ photon keV$^{-1}$ cm$^{-2}$ s$^{-1}$} \\
\multicolumn{2}{l}{\scriptsize $^{d}$ 10$^{-4}$ photon keV$^{-1}$ cm$^{-2}$ s$^{-1}$} \\
\multicolumn{2}{l}{\scriptsize $^{e}$ 10$^{-5}$ photon cm$^{-2}$ s$^{-1}$} \\
\end{tabular}
\hspace{-1 cm}
\label{mzWAfit}
\end{minipage}
\end{table}

We note that the model still includes a significant amount of cold absorption as a free parameter in the fit and a scattered power-law (similar to that detailed in Section~\ref{3-3}). There is a considerable decrease in the cold column density between 2001 and 2003 implying that it is not associated with the large scale galactic environment of Mrk 6. The drop in the cold column density results in the scattered power-law becoming an apparently more important component in the model fit to the 2003 observations. Despite this, the scattered power-law only contains $\sim$2.5\% of the flux of the underlying continuum in the fits to the 2001 data, an extremely similar value to that found from the 2003 \xmm ~observations.

\subsection{The origin of the complex absorption in Mrk 6}
\label{4-2}

Despite the progress made on the characterisation of the absorption in Mrk 6 it remains difficult to make any concrete statement on the physical origin of the absorption. The two models discussed here imply very different origins. The partial covering model suggests that the origin of the absorption arises from material in the broad line region, or closer, and that the absorbing column changes considerably (presumably through clouds moving across the line of sight) without significantly changing the partial covering fractions. Given the complexity of the partial covering model as it is, it is not possible to draw any clear conclusion on whether or not the partial covering material is genuinely cold from the data. However, under the assumption the partial covering material is not significantly ionised then the material must reside a considerable distance from the ionising source (the absorption profile from material with an ionisation parameter of $\la$1 is not considerably different from that of strictly cold material, and given an ionising luminosity of 10$^{43}$ erg s$^{-1}$ and a density of 10$^{12}$ cm$^{-3}$ results in an a cloud distance $>$100R$_{s}$). This does not rule out absorption in a wind from the accretion disk, particularly given the lack of tight constraints on the possible ionisation states of the partial covering media. 

Similarly, possible origins for the warm absorber interpretation include an ionised skin on the torus, warm gas in the vicinity of the broad line clouds or a disk wind, however in this instance the ionisation state can give us a handle on how plausible an origin in each of these regions is. For a given distance, ionising luminosity and measured ionisation parameter we can calculate the density of the absorbing material. Examination of these density values in comparison to the densities expected for the material in these regions can help us locate the absorbing material (we assume a 10$^{8}$\Msun black hole, an ionising luminosity of 10$^{43}$ erg s$^{-1}$ and distances of 60R$_{s}$, 10 light-days and 0.5pc for the disk wind, BLR and the inner edge of the torus respectively). Locating the absorption in a torus skin results in densities of $\sim$10$^{3\to5}$ cm$^{-3}$ which, at the upper limit, is consistent with the values calculated by Risaliti, Elvis and Nicastro (2002) for clouds in the torus with column densities of the order of 10$^{23}$ cm$^{-2}$. Locating the ionised material in the BLR (at $\sim$10 light-days; Peterson 1993) results in densities of $\sim$10$^{6\to7}$ cm$^{-3}$, considerably lower than the densities inferred for several AGN from ultraviolet observations ({\it e.g.} Gaskell \& Sparke 1986). Placing the absorption in a disk wind close to the ionising source results in densities of $\sim$10$^{9}$ cm$^{-3}$ which are in excellent agreement with the disk wind simulations by Proga \& Kallman (2004). Given the somewhat low values implied for a location at distances appropriate with the torus and BLR we favour a location in a disk wind extending on scales of tens to hundreds of Schwartzchild radii. In such a scenario the `double warm absorber' seen in the 2001 \xmm ~observation is easily explained as a line of sight which passes through two regions of the wind, one located somewhat closer to the ionising source (and hence more ionised) than the other. The apparent variability of the column density is then easily explained as an effect of the clumpy, outflowing turbulent nature of the wind which, given the possible range of outflow, radial \& turbulent velocities, may show absorption variability on timescales from hours upwards. The two observation of Mrk 6 discussed here are each insufficiently long to probe absorption variation on timescales longer than $\sim$4 hours making future long, or monitoring, observations of this source important to place any constraints on the nature of the wind.

\section{Conclusions}
\label{5}

A key feature of AGN unification schemes is the capability of absorbing material to change the appearance of a the central engine. To this end is it essential to understand the absorption present in these systems both to confirm or refute the standard model and to allow us to make conclusions regarding the central engine and the primary emission regions in AGN. 

In this paper we have presented an analysis of the most recent \xmm ~observation of Mrk 6. In particular, we test the hypothesis that the absorption can be well characterised by a warm absorption model that offers a more physical and testable description of the X-ray absorption than the standard, rather ad-hoc, partial covering model. We find that the simple warm absorption model, with only a single line of sight, provides an equally good statistical representation of the \xmm ~CCD data as a partial covering model. Furthermore, we find that once the RGS data are included in the spectral fitting, the simple warm absorber model provided a very good fit to the data, without adding any additional components to the model, contrasting with the partial covering model which requires the addition of either a low metalicity thermal plasma or blackbody emission in order to provide a similar quality fit. 

Further support for the presence of warm absorption comes when the previous \xmm ~observation is considered. The absorbing column density has varied considerably between the two observations. Fitting the partial covering model model to both sets of spectra does provide a reasonable fit and can explain the N$_{H}$ changes but, crucially, the covering fractions of the two dominant line of sight in the model have remained the same. To put this in context, this implies a situation in which absorption in moving clumpy material relatively close to the source results in two distinct lines of sight to the source, both with separate absorption that varies considerably without resulting in any appreciable change in the covering fractions involved. By contrast the warm absorber model needs only a second, higher column density but higher ionisation absorber to be present along the line of sight during the previous \xmm ~observation to explain the absorption variability. The presence of a second `phase' of the warm absorption is simple to interpret; an origin for the absorbing material in an ionised, outflowing, clumpy accretion disk wind or skin of the molecular torus provide a natural explanation for the absorption variability. The densities implied by the ionisation state of both the original warm absorber and the second, higher column density, higher ionisation, absorber are in good agreement with the densities found from simulations of these regions. Furthermore, the presence of several distinct ionisation `phases' is in good agreement with the results from the high resolution grating X-ray spectra of most bright type-1 AGN. Observations of these sources often display complex absorption with several distinct, clearly identifiable, ionization phases ({\it e.g.} NGC 3783 - Netzer \etal 2003, Behar \etal 2003, NGC 3516 - Turner \etal 2005, etc.).

Based on this evidence, coupled with the relatively arbitrary nature of the partial covering model, we strongly favour the interpretation of this absorption as originating in ionised clumpy material, probably resulting from an accretion disk wind. 

\section{Acknowledgments}

The authors wish to thank Chris Done for useful discussion and comments and the \xmm ~team for the hard work and dedication that underlies the success of the mission. NJS gratefully acknowledges financial support from PPARC and, with REG, also acknowledges the support of NASA grant NAG5-9902. This work is based on observations obtained with \xmm, an ESA science mission with instruments and contributions directly funded by ESA Member States and the USA (NASA). This research has made extensive use of NASA's Astrophysics Data System Abstract Service.

\end{document}